\documentclass[pre, aps, showpacs, superscriptaddress, twocolumn, longbibliography, groupedaddress]{revtex4-2}

\usepackage{hyperref}
\hypersetup{
     colorlinks=true,
     linkcolor=magenta,
     filecolor=blue,
     citecolor=blue,      
     urlcolor=cyan,
     }
\usepackage{color}
\usepackage[usenames,dvipsnames, table]{xcolor}
\usepackage{amsmath,amsthm,amssymb}
\usepackage{graphicx}
\usepackage{float}
\usepackage{bm}
\usepackage{mathrsfs}
\usepackage{multirow}
\usepackage{pbox}
\usepackage{verbatim}
\usepackage{braket}
\usepackage{mathtools}
\usepackage{bm}
\usepackage{mathtools}
\usepackage{tabstackengine}
\usepackage{enumerate}   
\usepackage{wasysym}
\usepackage{dsfont}
\usepackage{soul}
\usepackage{tabularx}
\usepackage{booktabs}
\usepackage{colortbl}
\usepackage{caption}
\usepackage{subcaption}
\usepackage{adjustbox}
\usepackage{tikz}
\usetikzlibrary{positioning,fit}
\usepackage{amsfonts}
\usepackage{booktabs}
\usepackage{siunitx}
\usepackage{pifont}

\stackMath

\newcommand{\er}[1]{Eq.~\eqref{#1}}
\newcommand{\ers}[2]{Eqs.~(\ref{#1}-\ref{#2})}

\newcommand{\Er}[1]{Equation~\eqref{#1}}

\newcommand{\WW}{\mathbb W}
\newcommand{\D}{\mathcal D}
\newcommand{\RR}{\mathbb R}
\newcommand{\R}{\mathcal R}
\newcommand{\n}{{\bf n}}
\newcommand{\stst}{| - \rangle}
\newcommand{\DKW}{\D_{\text{KW}}}
\newcommand{\ZZ}{\mathbb Z}
\newcommand{\DZZZ}{\D_{\text{ZZZ}}}
\newcommand{\DZXZ}{\D_{\text{ZXZ}}}
\newcommand{\id}{{\mathbb I}}

\newcommand{\beq}{\begin{equation}}
\newcommand{\eeq}{\end{equation}}

\begin{document}  

\title{Multicriticality in stochastic dynamics protected by self-duality}

\author{Konstantinos Sfairopoulos}
\affiliation{School of Physics and Astronomy, University of Nottingham, Nottingham, NG7 2RD, UK}
\affiliation{Centre for the Mathematics and Theoretical Physics of Quantum Non-Equilibrium Systems,
University of Nottingham, Nottingham, NG7 2RD, UK}
\author{Luke Causer}
\affiliation{School of Physics and Astronomy, University of Nottingham, Nottingham, NG7 2RD, UK}
\affiliation{Centre for the Mathematics and Theoretical Physics of Quantum Non-Equilibrium Systems,
University of Nottingham, Nottingham, NG7 2RD, UK}
\author{Juan P. Garrahan}
\affiliation{School of Physics and Astronomy, University of Nottingham, Nottingham, NG7 2RD, UK}
\affiliation{Centre for the Mathematics and Theoretical Physics of Quantum Non-Equilibrium Systems,
University of Nottingham, Nottingham, NG7 2RD, UK}

\begin{abstract}
    
We study the dynamical large deviations (LD) of a class of one-dimensional kinetically constrained models whose (tilted) generators can be mapped into themselves via duality transformations. We consider four representative models in detail: the domain-wall (DW) Fredrickson-Andersen (FA), the DW East, the ZZZ-FA, and the XOR-FA models. Using numerical tensor networks, we build the LD phase diagrams of these models in terms of the softness of the constraint and the counting field conjugate to the dynamical activity. In all cases, we find distinct dynamical phases separated by phase transitions along the self-dual lines, revealing the presence of multi-critical points that delimit first-order from continuous active-inactive transitions. We discuss connections to supersymmetry and possible extensions to higher spin and space dimensions. 
\end{abstract}

\maketitle

\noindent
{\bf \em Introduction.} This paper is about using ideas and methods most often associated with quantum many-body systems in the study of classical stochastic systems. When dynamics is subject to local constraints, such as the steric constraints of exclusion processes \cite{blythe2007nonequilibrium,mallick2015the-exclusion}, or the dynamical rules of kinetically constrained models (KCMs) \cite{fredrickson1984kinetic,jackle1991a-hierarchically,ritort2003glassy,garrahan2011kinetically,2024_Hartarsky}, classical stochastic systems can display complex correlated dynamics. A well-established approach to study such systems is that of dynamical large deviations (LDs), also known as the ``$s$-ensemble'' or ``thermodynamics of trajectories'', which aims to quantify the statistical properties of trajectory ensembles via LD methods \cite{merolle2005space-time,lecomte2005chaotic,garrahan2007dynamical,lecomte2007thermodynamic,hedges2009dynamic,garrahan2009first-order,touchette2009the-large,garrahan2018aspects,jack2020ergodicity,burenev2025an-introduction}. The statistics of dynamical (i.e., trajectory) observables is encoded in deformations of the operator that generates the stochastic dynamics, and since these deformed generators are often equivalent to quantum Hamiltonians, techniques for quantum systems, such as variational tensor networks
\cite{verstraete2008matrix,schollwock2011the-density-matrix,orus2014a-practical,silvi2019the-tensor,okunishi2022developments,banuls2023tensor}, are proving fruitful also in this classical context \cite{gorissen2009density-matrix,gorissen2012current,garrahan2016classical,banuls2019using,helms2019dynamical,helms2020dynamical,causer2020dynamics,causer2022finite,strand2022using,causer2023optimal,zima2025chemical}. 

Specifically, here we address two related problems in classical stochastic systems, the presence of \textit{ dualities} in dynamical generators and \textit{multicriticality} in the dynamics. The first of these issues connects to the current interest in quantum dualities 
\cite{aasen2016topological,ji2020categorical,aasen2020topological,lootens2023dualities,2024_Lootens,li2023noninvertible,seiberg2024majorana,seiberg2024non-invertible,gorantla2024tensor,chen2024sequential,lootens2024entanglement,obrien2025local}
from the perspective of generalized symmetries~\cite{gaiotto2015generalized,yoshida2016topological,mcgreevy2023generalized,cordova2022snowmass,bhardwaj2023lectures,schafer-nameki2023ictp,bhardwaj2024categorical,bhardwaj2025lattice}. The second issue connects to the recent interest in identifying multicritical behaviour at the level of LDs, both in the context of mean-field models \cite{tapias2024bringing} and of diffusive systems described by macroscopic fluctuation theory \cite{agranov2023tricritical}.

We study a class of one-dimensional KCMs with ``soft'' constraints \cite{buhot2001crossover,elmatad2010finite-temperature,elmatad2013space-time} which, at the level of LDs,
have a \textit{duality}: in each of these models, under a (generally) non-invertible operation, the \textit{tilted} generator that encodes trajectory statistics \cite{touchette2009the-large,garrahan2018aspects,jack2020ergodicity,burenev2025an-introduction} transforms to itself at different values of its parameters. We show that all models in this class have LD transitions between active and inactive dynamical phases, with the transitions located on the \textit{self-dual line} where the duality mapping leaves the tilted generator invariant. Furthermore, we show that these models display multicritical behaviour, with their active-inactive transition changing character, from first-order to continuous, at a multicritical point on the self-dual line. To our knowledge, this provides the first microscopic example of multicriticality in dynamical LDs.

\begin{figure*}[t]    
    \centering
    \includegraphics[width=\textwidth]{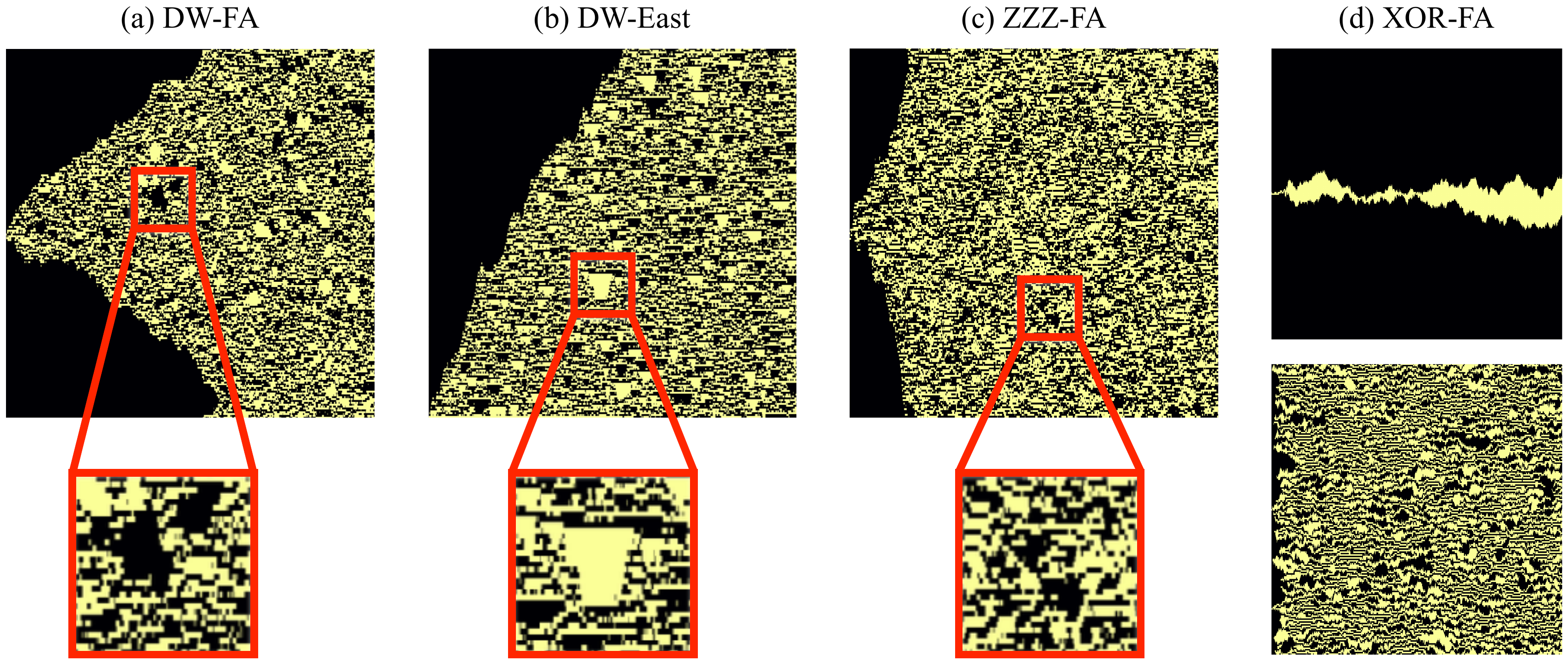}
    \caption{
    \textbf{Relaxation dynamics.} 
    (a) Typical relaxation trajectory for the DW-FA model for $\epsilon=0$, starting from a single spin up (yellow) in a background lattice of down spins (black). The system size is $N = 200$ (with OBC) and the overall time $t = 200$. The inset illustrates the fluctuations in the dynamics in an equilibrated space-time region of size $32 \times 32$.
    (b-c) Same for the DW-East and ZZZ-FA models. 
    (d) Relaxation trajectory with two DWs in the XOR-FA model at $\epsilon=0$ (top), and at $\epsilon=0.05$ (bottom).
    }
    \label{fig:trajectories}
\end{figure*}

\smallskip

\noindent
{\bf \em Models.} We consider one-dimensional lattice models with a binary variable or spin, $n_i = 0$ or $1$, on each lattice site $i$ ($i = 1, \ldots, N$), with dynamics governed by a continuous-time Markov generator of the form
\begin{equation}{\label{eq:W}}
    \WW_{\epsilon} 
    = 
    \sum_{i=1}^N  \left(f_i+ \epsilon \right) \left( X_i - 1 \right),
\end{equation}
with $X_i$ the Pauli $X$-matrix on site $i$. The \textit{kinetic constraint} $f_i$ is a diagonal operator that depends on the neighbours of $i$. The parameter $\epsilon \geq 0$ tunes the strength of the constraint: for $\epsilon=0$, \er{eq:W} it describes a KCM with a \textit{hard constraint}, meaning that flips can occur only when the neighbours of $i$ are such that $f_i > 0$; when $\epsilon > 0$ (\textit{soft constraint}) this condition is relaxed \cite{buhot2001crossover,elmatad2010finite-temperature,elmatad2013space-time}. In the limit of very large $\epsilon$ the constraint is irrelevant and the dynamics of the spins becomes non-interacting. The generator \eqref{eq:W} is bi-stochastic, meaning that the stationary state is the flat state (or ``infinite temperature'' state), 
$\ket{P_{\text{ss}}} = 2^{-N} \stst = 2^{-N} \sum_{\{\n\}} | \n \rangle$, 
where all configurations $\n = (n_1, \ldots, n_N)$ are equally probable
\footnote{
    The flat state is the unique stationary state if the dynamics is irreducible. This may not be the case in the hard constraint case, where the flat state in each ergodic component is stationary. For $\epsilon > 0$, ergodicity is restored, up to global symmetries. 
}. As such, $\WW_{\epsilon}$ is also Hermitian and can alternatively be considered (minus) the Hamiltonian of a spin-$1/2$ quantum chain. 

We will consider four cases for the constraint $f_i$. These are
\begin{align}
    f^{\text{DW-FA}}_i 
    &= 
    1 - \frac{1}{2} \left(Z_{i-2}Z_{i-1} + Z_{i+1}Z_{i+2}\right),
    \label{eq:DWFA}  
    \\
    f^{\text{DW-East}}_i 
    &= 
    1 - Z_{i-2}Z_{i-1},  
    \label{eq:DWEast}  
    \\
    f^{\text{ZZZ-FA}}_i 
    &= 
    1  - \frac{1}{2} \left(Z_{i-3}Z_{i-2}Z_{i-1} + Z_{i+1}Z_{i+2}Z_{i+3}\right),  
    \label{eq:ZZZFA}  
    \\
    f^{\text{XOR-FA}}_i 
    &= 
    1  - Z_{i-1}Z_{i+1},  
    \label{eq:XORFA}  
\end{align}
where $Z_i$ the Pauli $Z$-matrix on site $i = 1$ taking values $\pm 1$ for an up/down spin at site $i$. 
\Er{eq:W} with the constraint \eqref{eq:DWFA} is a domain wall (DW) version of the much-studied Fredrickson-Andersen (FA) model \cite{fredrickson1984kinetic,ritort2003glassy,garrahan2011kinetically}, such that in the hard case ($\epsilon = 0$) a site can only flip if there is at least one domain wall to its left or right (compared to the standard FA model where a site can flip if at least one neighbour is in the up state). 
\Er{eq:W} with constraint \eqref{eq:DWEast} is the DW version of the East KCM \cite{jackle1991a-hierarchically,ritort2003glassy,garrahan2011kinetically,2024_Hartarsky}, where the hard constraint allows flips only if there is a DW to the left of the site. \Er{eq:W} with constraint \eqref{eq:XORFA} defines a soft version of the XOR-FA model \cite{causer2020dynamics}. Constraint \eqref{eq:ZZZFA} is similar to the DW-FA ones, but flips only occur if at least one of the triplets of the spins to the left and right of the site have negative parity. 

Figure 1 shows typical trajectories of the dynamics for the four models, \ers{eq:W}{eq:XORFA}, starting from an initial state with a single up spin. Figures 1(a-c) correspond to the DW-FA, DW-East, and ZZZ-FA models with a hard constraint, $\epsilon=0$. In all cases, we observe that the relaxation towards the stationary state $\stst$ is via an outward growth of an active/equilibrated region, similar to what is seen in other KCMs \cite{blondel2013front,katira2016pre-transition}. The insets indicate that in equilibrium, there are noticeable space-time fluctuations indicative of dynamical heterogeneity \cite{garrahan2018aspects}. These are typical indicators of the coexistence of active and inactive phases \cite{merolle2005space-time,
garrahan2007dynamical,garrahan2018aspects}. Figure 1(d) shows the trajectories of the XOR-FA for $\epsilon =0$ (top) and $\epsilon = 0.05$ (bottom): in the hard constraint case, domain walls are conserved, but for any $\epsilon>0$ the system relaxes to $\stst$.

\begin{figure*}[t]
    \centering
    \includegraphics[width=\linewidth, height=7.8cm]{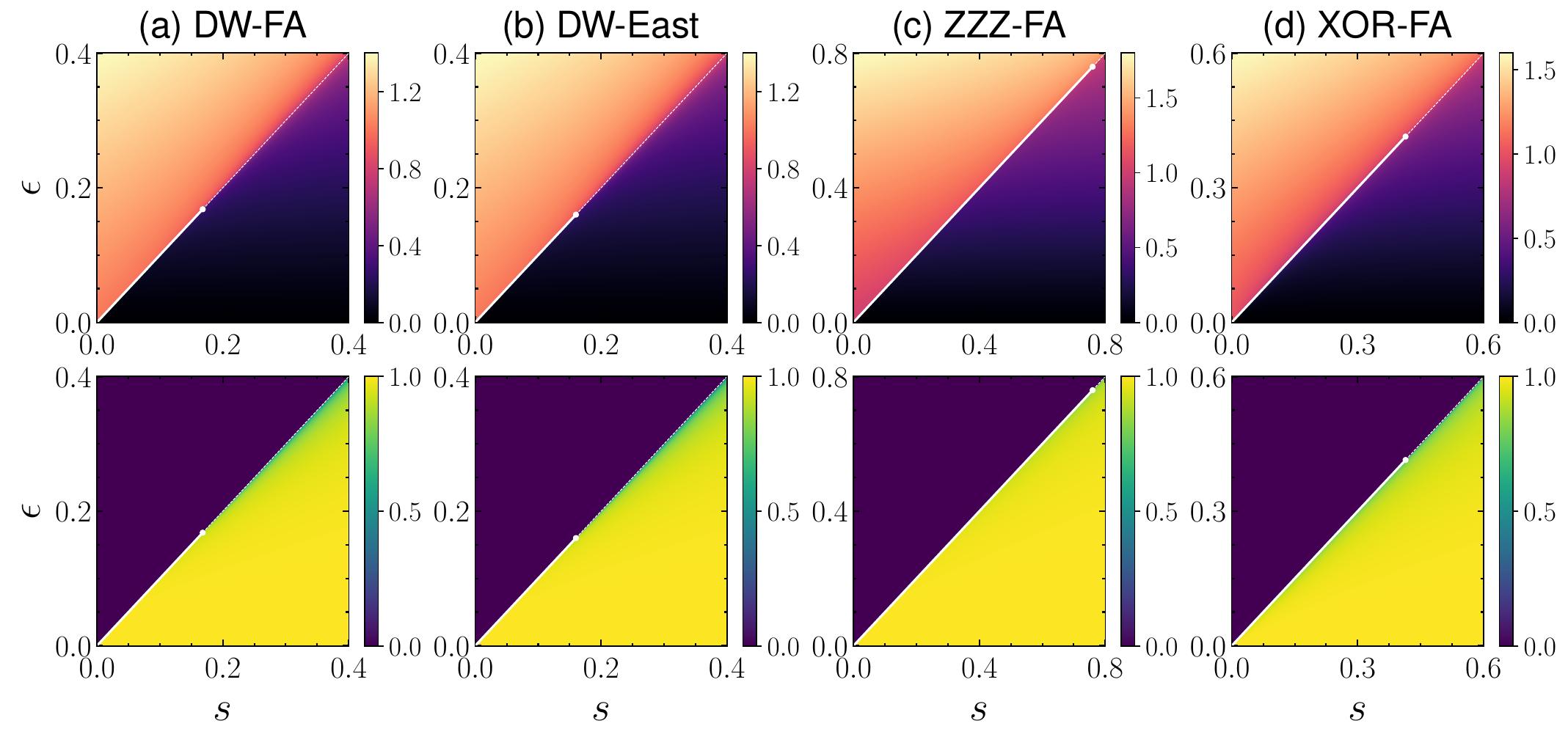}
    \caption{ 
    \textbf{Large deviation phase diagrams.} 
    (a) LD phase diagram of the DW-FA model obtained by VUMPS. The top panel shows the average time-integrated escape rate (per unit time and size) as a function of $\epsilon$ and $s$. The bottom panel shows the average magnetization for the leading eigenstate of the tilted generator. The first-order transition line (solid-white) is separated from the continuous transition line (dashed-white), by a tricritical point at $\epsilon_{\text{TC}} = s_{\text{TC}} \approx 0.168$. 
    (b) Same for the DW-East, with $\epsilon_{\text{TC}} = s_{\text{TC}} \approx 0.16$.
    (c) Same for the ZZZ-FA model, with $\epsilon_{\text{TC}} = s_{\text{TC}} \approx 0.76$.
    (d) Same for the XOR-FA model, with $\epsilon_{\text{TC}} = s_{\text{TC}} = \sqrt{2} - 1$.
    }
    \label{fig:phase_diagram}
\end{figure*}

\smallskip

\noindent
{\bf \em Tilted generators.} We study the dynamics of the above models using the by-now standard tools of dynamical LDs \cite{touchette2009the-large,garrahan2018aspects,jack2020ergodicity,burenev2025an-introduction}. As a trajectory observable, we consider the time integral of the escape rate, 
\beq
    \R(\omega_t) 
    = 
    \int_{0}^{t} dt' 
    \langle \n_{t'}(\omega_t) | 
        \RR 
    | \n_{t'}(\omega_t) \rangle 
    \, ,
    \label{eq:R}
\eeq
where $\omega_t$ indicates a stochastic trajectory with $t' \in [0, t]$, and $\n_{t'}(\omega_t)$ the configuration of the system in that trajectory at time $t'$. The diagonal operator $\RR$ is the matrix of escape rates and is given by, cf.\ \er{eq:W}, 
\begin{equation}
    \label{eq:RR}
    \RR 
    = 
    \sum_{i=1}^N  \left(f_i+ \epsilon \right).
\end{equation}
The observable $\R$ is directly related to the dynamical activity (number of configuration changes), and thus quantifies the overall amount of dynamics in a trajectory, see \cite{garrahan2009first-order,garrahan2018aspects,maes2020frenesy:}.

The moment generating function (MGF) of $\R$ takes the form of a partition sum over trajectories
\cite{touchette2009the-large,garrahan2018aspects,jack2020ergodicity,burenev2025an-introduction} 
\beq
    Z_{t}(s) = 
    \sum_{\{\omega_t\}} 
    \pi(\omega_{t}) \, 
    e^{-s \R(\omega_t)} 
    = 
    \langle - | e^{t \WW_{\epsilon,s}} \ket{P_{\text{ss}}},
    \label{eq:Zs}
\eeq
where $\pi(\omega_t)$ the probability of trajectory $\omega_t$ under the dynamics. Note that in \er{eq:Zs} we assume that the dynamics starts from an equilibrium state (in contrast to the example trajectories of Fig.~\ref{fig:trajectories}). \Er{eq:Zs} defines the tilted generator
\cite{touchette2009the-large,garrahan2018aspects,jack2020ergodicity,burenev2025an-introduction} 
\begin{equation}{\label{eq:Ws}}
    \WW_{\epsilon,s} 
    = 
    \sum_{i=1}^N  \left(f_i+ \epsilon \right) \left[ X_i - 
    (1+s) \right],
\end{equation}
which encodes the statistics of the observable $\R$ for all times. The tilted generators for the four models, obtained by combining \er{eq:Ws} with \ers{eq:DWFA}{eq:XORFA}, are the key objects of our study. In the following, we will only consider the case $s \geq 0$.

\begin{figure*}[t]
    \centering
    \includegraphics[width=\linewidth]{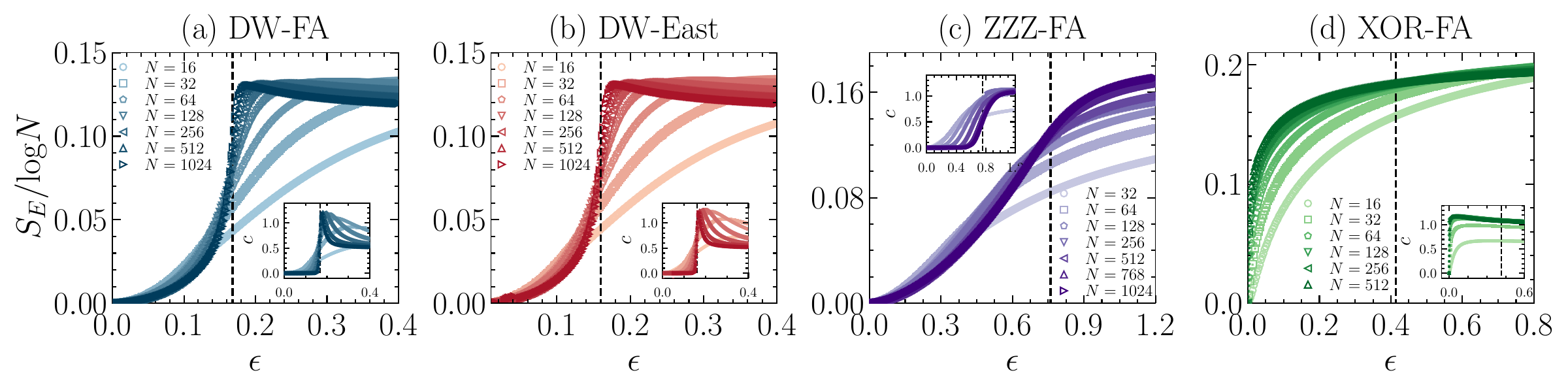}
    \caption{
    \textbf{Locating the tricritical point.} 
    (a) Estimate of the tricritical point in the DW-FA. Main panel is the scaled bipartite entanglement entropy, $S_{E} / \log N$, along the self-dual line. Insets is the central charge $c$. The black dashed line indicates the location of the tricritical point estimated from the peak of the entanglement susceptibility $\chi_{E} = \frac{1}{\log N} dS_{E} / ds$.
    (b) Same for the DW-East. 
    (c) Same for the ZZZ-FA, where the location of the multicritical point is found from the derivative of the central charge. 
    (d) Same for the XOR-FA.
    }
    \label{fig: entanglement}
\end{figure*}

\smallskip

\noindent
{\bf \em Invertible symmetries, non-invertible symmetries, and dynamical phases.} We now consider the properties of each of the four models. While all these are new stochastic models (as far as we know), in some cases, their tilted generators have been studied as Hamiltonians of closely related quantum spin chains. By combining existing results (which we confirm numerically) with new ones, we classify their dynamics.

\smallskip

{\em (i) DW-FA model.} The tilted generator $\WW^{\text{DW-FA}}_{\epsilon,s}$, obtained from \er{eq:Ws} with constraint \eqref{eq:DWFA}, is by far the best understood of our models when seen as a quantum spin chain. Ref.~\cite{obrien2018lattice} studied it in the context of supersymmetric (SUSY) quantum models on the lattice \cite{nicolai1976supersymmetry}. It has also been considered among others in the context of conformal field theories (CFTs) \cite{zou2018conformal,zou2020emergence,zou2021universal,kuo2022decoding,vardhan2023petz,cogburn2024cft-and-lattice,koyluoglu2024measuring}, of the false vacuum decay \cite{milsted2022collisions}, of novel phases under measurements \cite{patil2024highly}, and of state preparation in quantum computers \cite{roy2024efficient}. A variation of the model \cite{sannomiya2019supersymmetry} and its antiferromagnetic version \cite{slagle2021microscopic} have also been studied.

The tilted generator $\WW^{\text{DW-FA}}_{\epsilon,s}$ has a $\ZZ_2$ symmetry under the action of $\eta = \prod_i X_i$ (global spin-flip), and is translationally invariant (with PBC). There is also a Kramers-Wannier (KW) duality: a \textit{non-invertible} operator $\DKW$ (for its explicit form see e.g.\ Refs.~\cite{seiberg2024majorana,seiberg2024non-invertible,gorantla2024tensor}) acts on the terms of the Hamiltonian as 
\begin{align}
    \DKW X_i 
    &= Z_{i-1}Z_i \DKW,
    \\
    \DKW Z_i Z_{i+1} 
    &= X_i \DKW,
\end{align}
so that 
\begin{equation}
    \label{eq:KWFA}
    \DKW \WW^{\text{DW-FA}}_{\epsilon,s} 
    = 
    \WW^{\text{DW-FA}}_{s,\epsilon} \DKW. 
\end{equation}
On the self-dual line, $\epsilon = s$, the transformation $\DKW$ commutes with the tilted generator. The duality \eqref{eq:KWFA} connects the active phase of a softer DW-FA model with the inactive phase of a harder one. 

The dynamical phases encoded by $\WW^{\text{DW-FA}}_{\epsilon,s}$ can be inferred from the results of Ref.~\cite{obrien2018lattice} for the Hamiltonian (up to a constant) ${\mathbb H}(\epsilon, s) = -\WW^{\text{DW-FA}}_{\epsilon,s}$. Based on a renormalization group (RG) argument, it was shown that, for large $\epsilon = s$, the model flows to the stable Ising fixed point and is gapless. In contrast, for smaller $\epsilon = s$, it flows to the stable free fermion fixed point $\epsilon = s = -1$ of the three-body interactions \cite{fendley2019free} with a dynamical critical exponent $z=3/2$, and is gapped. These two regimes are separated by the unstable Ising tricritical \cite{blume1971ising,nienhuis1979first-} fixed point at $\epsilon_{\text{TC}} = s_{\text{TC}} \approx 0.168$ \cite{obrien2018lattice}.

These results translate to the dynamics of the DW-FA model as follows. For $\epsilon > 0$ and $s=0$ dynamics is active and ergodic, cf.\ Fig.~\ref{fig:trajectories}. In turn, for a hard constraint ($\epsilon = 0$), we expect to have an LD first-order transition at $s=0^+$ to an inactive phase, in analogy to Ref.~\cite{garrahan2007dynamical}. This means that the self-dual line of the DW-FA model separates an active phase at $\epsilon > s$ from an inactive one at $\epsilon < s$. This transition is first-order for $\epsilon < \epsilon_{\text{TC}}$ and continuous for $\epsilon > \epsilon_{\text{TC}}$. 

We can understand the properties of its dynamical phases from their transformation rules under the symmetries. When both $\epsilon$ and $s$ are large, the operator $-\WW^{\text{DW-FA}}_{\epsilon,s}$ is essentially the Hamiltonian of the transverse-field Ising model. In that case, there are two phases. For $s > \epsilon \gg 1$, the classical ground states $\ket{0\cdots 0} \equiv \ket{\mathbf{0}} $ and $\ket{1\cdots 1} \equiv \ket{\mathbf{1}}$ spontaneously break the global symmetry. For $\epsilon > s \gg 1$ we are in the symmetric phase with ground state $\stst$. Under the symmetries, these states transform as 
\begin{align}
    \eta \ket{\mathbf{0}} 
    &= 
    \ket{\mathbf{1}}, 
    \;
    &
    \eta \ket{\mathbf{1}} 
    &= \ket{\mathbf{0}}, 
    \;
    & 
    \eta \stst 
    &= \stst 
    \nonumber
    \\
    \DKW \ket{\mathbf{0}} 
    &= \stst, 
    \;
    &
    \DKW \ket{\mathbf{1}} 
    &= \stst, 
    \;
    &
    \DKW \stst 
    &= \ket{\mathbf{0}} + \ket{\mathbf{1}} 
    \nonumber
\end{align}
On the self-dual line, the phase transition at $\epsilon_{\text{TC}}$ corresponds to the spontaneous breaking of the non-invertible symmetry \cite{seiberg2024majorana,seiberg2024non-invertible}. 

Based on various Lieb-Schultz-Mattis (LSM) type theorems formulated in recent years \cite{levin2020constraints,seiberg2024majorana,seiberg2024non-invertible}, self-dual Ising chains were shown to be gapless or degenerate. As a result, the first-order phase transition of the self-dual line of the DW-FA model carries a three-fold degeneracy, which was studied in Refs.~\cite{obrien2018lattice,seiberg2024non-invertible} and proven in \cite{gorantla2024duality-preserving}. 

These expectations are confirmed by numerical tensor network simulations. In Fig.~\ref{fig:phase_diagram}(a), we show the LD phase diagram of the DW-FA model obtained by estimating the leading eigenvector of $\WW^{\text{DW-FA}}_{\epsilon,s}$ using 
variational uniform matrix product states (VUMPS) \cite{zauner-stauber2018variational}. The top panel shows the average escape rate, corresponding to the mean \er{eq:R} per unit time, while the bottom panel shows the average absolute magnetization. We can see that the self-dual line separates two phases of distinct activity. In order to better estimate the location of the singularities on the self-dual line, we use DMRG \cite{white1992density,itensor1,itensor2}. Figure~\ref{fig: entanglement}(a) shows the scaled bipartite entanglement entropy along the self-dual line, and the central charge, $c$, in the inset calculated for open boundary conditions following Ref.~\cite{2011_Xavier} from the underlying CFT \cite{1994_Holzhey,2004_Calabrese}. For $0 < \epsilon = s < \epsilon_{\text{TC}}$, we observe a gapped phase, a result consistent with the first-order transition line. For $\epsilon = s > \epsilon_{\text{TC}}$ we obtain a critical phase with central charge $c=1/2$, consistent with the Ising CFT. The tricritical Ising CFT governs the tricritical point \cite{di-francesco1997conformal}.

\smallskip

{\em (ii) DW-East model.} The tilted generator $\WW^{\text{DW-East}}_{\epsilon,s}$ is related, with the addition of an extra fine-tuned boundary term, to the interacting Kitaev chain, which has $\mathcal{N}=1$ SUSY, as studied in Ref.~\cite{miura2024interacting}. Via numerics and analytical bounds, that model was found to have tricritical behaviour in the Ising tricritical universality class \cite{miura2024emergent,sannomiya2024spontaneous}. While the parameters of $\WW^{\text{DW-East}}_{\epsilon,s}$ are different from that of the model in Refs.~\cite{miura2024interacting,miura2024emergent}, one can think of an enlarged parameter space where the two tricritical points are connected by a tricritical line, similar to Ref.~\cite{obrien2018lattice}. 

The DW-East model has the same global symmetry as the DW-FA model. Due to the one-directional nature of its constraint, see \er{eq:DWEast}, the KW duality for the DW-East \eqref{eq:KWFA} reads 
\begin{equation}
    \label{eq:KWEast}
    \DKW \WW^{\text{DW-East}}_{\epsilon,s} 
    = 
    \WW^{\text{DW-West}}_{s,\epsilon} \DKW, 
\end{equation}
where the DW-West model is the same model but space-inverted. A relation of the DW-East and DW-West models was recently used to explain the counting of the Nambu-Goldstone modes, based on an emergent chiral supersymmetry on the continuum, see Ref.~\cite{nakayama2025is-chiral}.

Figure~\ref{fig:phase_diagram}(b) shows that the LD phase diagram of the DW-East model is similar to that of the DW-FA. We estimate the location of the tricritical point to be $\epsilon_{\text{TC}} = s_{\text{TC}} \approx 0.16$, see Fig.~\ref{fig: entanglement}(b). As for the DW-FA model, in the critical phase at $\epsilon = s > \epsilon_{\text{TC}}$, we measure an Ising CFT central charge $c=1/2$, see inset of Fig.~\ref{fig: entanglement}(b).

\smallskip

{\em (iii) ZZZ-FA model.} As far as we are aware, the tilted generator $\WW^{\text{ZZZ-FA}}_{\epsilon,s}$ is new when considered as the Hamiltonian of a quantum spin chain. At the point $\epsilon=s\gg 1$, it corresponds to the three-spin Ising model 
governed by the 4-state Potts CFT \cite{penson1982phase,turban1982self-dual,maritan1984first-,igloi1986series,kolb1986conformal,alcaraz1987conformal,udupa2023weak}. Its self-duality in this limit was studied in Refs.~\cite{yan2024generalized,gorantla2024tensor}.

For system sizes $N = 3 \times k$ with $k$ integer, the ZZZ-FA model has a $\ZZ_2 \times \ZZ_2$ symmetry under the action of $\eta_1 = \prod_{i \notin 3 \mathbb{Z} + 1} X_i$ and $\eta_2 = \prod_{i \notin 3 \mathbb{Z} + 2} X_i$, with $\mathbb{Z}$ the integers. The non-invertible operator $\DZZZ$ transforms the terms of the generator as follows 
\begin{equation}
    \DZZZ Z_{i-1}Z_i Z_{i+1} = X_i \DZZZ,
\end{equation}
with $\DZZZ^2 = (1+\eta_1)(1+\eta_2)$ \cite{gorantla2024tensor}. 

Figure~\ref{fig:phase_diagram}(c) shows that, like the DW-FA and DW-East models, the ZZZ-FA model has active and inactive dynamical phases subdivided by the same line, $\epsilon = s$. For sizes $N = 3 \times k$, along this line, the ZZZ-FA model is gapless at large $\epsilon$, and gapped 
at small $\epsilon$ with coexistence of five ground states. We expect similar LSM-type arguments to hold, cf.\ Refs.~\cite{levin2020constraints,seiberg2024non-invertible}. For system sizes not a multiple of three we expect a two-state coexistence and no nontrivial global symmetries. The non-invertible operator in this case is expected to become invertible \cite{seo2024non-invertible,gorantla2024tensor}. The tricritical point is located around $\epsilon_{\text{TC}} = s_{\text{TC}} \approx 0.76$, see Fig.~\ref{fig: entanglement}(c). Note that the simulations for the ZZZ-FA model were much harder to converge numerically, and we believe that the thermodynamic limit has not been reached yet from the system sizes studied.
The critical phase for $\epsilon = s > \epsilon_{\text{TC}}$ is known to fall in the 4-state Potts universality class \cite{udupa2023weak} with central charge $c=1$, which we observe in our numerics, see Fig.~\ref{fig: entanglement}(c). It is unclear if any (and which) CFT characterizes the tricritical point
\footnote{
    Based on the exact diagonalization of small systems (not shown), we estimate a dynamical critical exponent $z=1$ at the tricritical point from the size scaling of the spectral gap. }.

\smallskip

{\em (iii) XOR-FA model.} 
The LDs of the hard ($\epsilon=0$) XOR-FA model were first studied in Ref.~\cite{causer2020dynamics}. Considered as a quantum Hamiltonian, the generator $\WW^{\text{XOR-FA}}_{\epsilon,s}$ for arbitrary coefficients was subsequently studied in the context of gapless topological phases \cite{verresen2021gapless}, through a web of dualities \cite{moradi2023topological,li2023noninvertible,ando2024a-journey,eck2023critical,eck2024from,cao2025global,seifnashri2025gauging,2025_Shao}, and at finite temperature \cite{tavares2023finite}.

The generator $\WW^{\text{XOR-FA}}_{\epsilon,s}$ has two global $\ZZ_2$ symmetries, $\eta_{\text{e}} = \prod_{\text{even}} X_i$ and $\eta_{\text{o}} = \prod_{\text{odd}} X_i$. It has a duality mapping \cite{zadnik2023slow,dober2024on-antiferromagnetic}
to the XYZ model which is integrable \cite{baxter2016exactly,fendley1989non-critical}.    
$\WW^{\text{XOR-FA}}_{\epsilon,s}$ has terms of the form $ZXZ$, $Z \id Z$, and $X$. If any one of these vanishes \cite{2025_Su}, then the generator becomes exactly solvable via a Jordan-Wigner transformation \cite{minami2016solvable,minami2017infinite,yanagihara2020exact}, and has additional dualities directly coming from quantum field theory (T-duality) \cite{pace2024lattice}.
Furthermore, when the coefficients of any of these two terms are the same
\footnote{
    In our notation this occurs when $\epsilon = s$, $s = -2$, or $\epsilon = -2$. The latter is unphysical for stochastics as it corresponds to non-positive generators.
}, 
it maps to the XXZ chain, which has SUSY for open boundary conditions (OBC) at a special point~\cite{fendley2003lattice,yang2004non-local,weston2017lattice}. With only the 
$Z X Z$ term \cite{raussendorf2003measurement-based,doherty2009identifying,son2012topological,chen2014symmetry-protected} 
there is a self-duality under 
an operator $\DZXZ$ \cite{seifnashri2024cluster,parayil-mana2024kennedy-tasaki}. 
Since 
\begin{equation}
    \DZXZ X_i = Z_{i-1} Z_{i+1} \DZXZ,  
\end{equation}
this $\text{Rep}(\mathcal{D}_8)$ symmetry \cite{seifnashri2024cluster} is also a symmetry for $\WW^{\text{XOR-FA}}_{\epsilon,s}$. 

For $\WW^{\text{XOR-FA}}_{\epsilon,s}$ and $s > \epsilon \gg 1$ there are four leading eigenstates, ($\ket{\uparrow \cdots \uparrow}$, $\ket{\downarrow \cdots \downarrow}$, $\ket{\uparrow \downarrow\cdots \uparrow \downarrow}$, $\ket{\downarrow \uparrow \cdots \downarrow \uparrow}$), while for $\epsilon > s \gg 1$ there is a single one given by the flat state. At  $\epsilon=0$ there is a first-order transition at $s=0^+$ \cite{lieb167exact-ferroelectric,lieb1967exact-ice,kohmoto1981hamiltonian} which for the stochastic XOR-FA is an active-inactive LD transition \cite{causer2020dynamics}, and the system is gapped (central charge $c=0$). The active and inactive phases are separated by the self-dual line, with $\epsilon=s < \sqrt{2} - 1$ being first-order (but with a diverging specific heat with exponent $\alpha = 1/2$ \cite{kohmoto1981hamiltonian}). For $\epsilon=s>\sqrt{2} - 1$ the transition is critical with continuously varying critical exponents \cite{baake1987operator,henkel1988conformal,yang1987modular,yang1987superconformal,bridgeman2015multiscale,obrien2015symmetry-respecting}. For $\epsilon = s \to \infty$ we expect a central charge $c=1$. Our numerics confirms these results, see Fig.~\ref{fig:phase_diagram}(d) and Fig.~\ref{fig: entanglement}(d). 

\smallskip

\noindent
{\bf \em Conclusions.} There are many more quantum spin models that can have a classical stochastic interpretation, regarding generators with single or multiple spin-flip stochastic dynamics, such as those of Refs.~\cite{obrien2018lattice,chatterjee2024quantum,sannomiya2016supersymmetry,sannomiya2019supersymmetry,2025_Rey}. 
What we did here for stochastic lattice models with binary variables can be extended to models with higher local state spaces with dualities \cite{wu1982the-potts,obrien2020self-dual,obrien2020not-a,seo2024non-invertible}. For example, for Potts spins with more than four states intermediate Berezinskii–Kosterlitz–Thouless transitions or critical regions might be found, cf.\ Ref.~\cite{ortiz2012dualities}. Furthermore, the approach followed for the 1D stochastic models of this work can be extended to higher-dimensional models with dualities where topologically ordered phases can be present \cite{kitaev2003fault-tolerant,kitaev2006anyons} or to deconfined quantum criticality \cite{2004_Senthil,2017_Wang,2023_Senthil}. We also note that in analogy to its quantum counterpart, the XOR-FA model has a symmetry-protected topological (SPT) phase \cite{verresen2021gapless}, which one can generalize to models with longer-ranged constraints leading to interesting connections to other SPT phases \cite{yoshida2016topological}, whose implications for classical stochastic dynamics are not fully understood \cite{2018_Dasbiswas,garrahan2024topological}.

\smallskip

\noindent
{\bf \em Acknowledgments.}
    We thank A. Foligno for discussions on CFTs.
    We acknowledge financial support from EPSRC Grant No.\ EP/V031201/1 
    and Leverhulme Trust Grant No.\ RPG-2024-112. 
    Simulations were performed using the University of Nottingham Augusta and Ada HPC clusters (funded by EPSRC Grant No.\ EP/T022108/1 and the HPC Midlands+ consortium). The DMRG calculations were performed with use of the Julia ITensor library \cite{itensor1,itensor2}.

\bibliography{bibliography-31032025}
\bibliographystyle{apsrev4-2}

\end{document}